# Resurfacing processes constrained by crater distribution on Ryugu

**Naofumi Takaki[1], Yuichiro Cho[1], Tomokatsu Morota[1,2], Eri Tatsumi[1,3], Rie Honda[4], Shingo Kameda[5], Yasuhiro Yokota[4,6], Naoya Sakatani[6], Toru Kouyama[7], Masahiko Hayakawa[6], Moe Matsuoka[6], Manabu Yamada[8], Chikatoshi Honda[9], Hidehiko Suzuki[10], Kazuo Yoshioka[1], Kazunori Ogawa[11,12], Hirotaka Sawada[6], Patrick Michel[13], Seiji Sugita[1,8]**

[1] The University of Tokyo, Tokyo 113-0033, Japan.

[2] Nagoya University, Nagoya 464-8601, Japan.

[3] Instituto de Astrofísica de Canarias (IAC), 38205 La Laguna, Tenerife, Spain.

[4] Kochi University, Kochi 780-8520, Japan.

[5] Rikkyo University, Tokyo 171-8501, Japan.

[6] Institute of Space and Astronautical Science (ISAS), Japan Aerospace Exploration Agency (JAXA), Sagamihara 252-5210, Japan.

[7] National Institute of Advanced Industrial Science and Technology, Tokyo 135-0064 Japan.

[8] Planetary Exploration Research Center, Chiba Institute of Technology, Narashino 275-0016, Japan.

[9] University of Aizu, Aizu-Wakamatsu 965-8580, Japan.

[10] Meiji University, Kawasaki 214-8571, Japan.

[11] Kobe University, Kobe 657-8501, Japan.
[12] JAXA Space Exploration Center (JSEC), Japan Aerospace Exploration Agency (JAXA), Sagamihara 252-5210, Japan.

[13] Université Côte d'Azur, Observatoire de la Côte d'Azur, CNRS, Laboratoire Lagrange, 06304 Nice Cedex 4, France.

Corresponding author: Naofumi Takaki (ntakaki@eps.s.u-tokyo.ac.jp)

**Abstract**

Understanding the geological modification processes on asteroids can provide information concerning their surface history. Images of small asteroids from spacecraft show a depletion in terms of smaller craters. Seismic shaking was considered to be responsible for crater erasure and the main driver modifying the geology of asteroids via regolith convection or the Brazil nut effect. However, a recent artificial impact experiment on the asteroid Ryugu by the Japanese Hayabusa2 mission revealed minimal seismic activity. To investigate whether a seismic shaking model can reproduce the observed crater record, the crater distribution on Ryugu was analyzed using crater production functions under cohesionless conditions. Crater retention ages were estimated as a function of crater diameter for Ryugu, Itokawa, Eros, and Bennu using the crater size-frequency distribution and crater production function estimated for those asteroids. We found that the power-law indices $a$ are inconsistent with diffusion processes (e.g., seismic shaking, $a = 2$). This result suggests that seismic shaking models based on diffusion equations cannot explain the crater distribution on small asteroids. Alternative processes include surface flows, possibly at the origin of geomorphological and spectral features of Ryugu. We demonstrate that the vertical mixing of material at depths shallower than 1 m occurs over $10^3$–$10^5$ yr by cratering and obliteration. The young surface age of Ryugu is consistent with the slow space weathering that results from cratering, as suggested in previous studies. The timescale ($10^4$–$10^6$ yr) required for resurfacing at depths of 2–4 m can be compared with the cosmic-ray exposure ages of returned samples to constrain the distribution of impactors that collide with Ryugu.

**1 Introduction**

The crater distribution of small asteroids is depleted in terms of smaller craters. The relative size-frequency distribution of craters on asteroid 433 Eros shows a deficit of craters smaller than 100 m in diameter (Chapman et al., 2002). A similar depletion was observed on asteroid 25143 Itokawa, 2867 Šteins, and 4179 Toutatis (Hirata et al., 2009; Besse et al., 2012; Barucci et al., 2015). A deficit of small impactors producing craters smaller than 100 m in diameter is not expected based on our current understanding of the size distribution of asteroid populations and therefore such depletions are interpreted as evidence of resurfacing processes on these asteroids. Many types of resurfacing processes have been proposed to account for the lack of small craters: seismic shaking induced by impacts (Richardson et al., 2005; Michel et al., 2009), granular convection (Miyamoto et al., 2007; Yamada et al., 2016), and dust levitation (Hirata and Miyamoto, 2012). Additionally, the armoring effect, in which the energy of an impact is consumed by the fragmentation of boulder located at the impact point rather than forming a crater, was proposed as a possible cause of the crater depletion on Itokawa (Tatsumi & Sugita, 2018; Barnouin et al., 2019; Daly et al., 2019). Global observations of asteroid 162173 Ryugu also demonstrate a pronounced depletion in craters smaller than 100 m (Sugita et al., 2019; Cho et al., 2021).

In previous studies, seismic shaking was assumed to be the main process by which the surface of an asteroid is modified. Geologic processes that move and sort boulders on asteroids, such as the Brazil nut effect (e.g., Matsumura et al., 2014) and regolith convection (Miyamoto et al., 2007; Yamada et al., 2016), also assume the occurrence of seismic shaking as a driver. Seismic shaking was previously modeled to account for the deviation in the observed crater distributions of craters smaller than 100 m on Eros and Itokawa from the crater production function on these asteroids (Michel et al., 2009; Richardson et al., 2020). Model parameters, such as the diffusion coefficient, the thickness of the regolith layers, the quality factor $Q$ for seismic waves, and the large cohesion coefficient (5 MPa for Richardson et al. (2020) and 18 MPa for Michel et al. (2009)) were assumed in the models. However, the Small Carry-on Impactor (SCI) experiment performed by the Japanese mission Hayabusa2, which produced an artificial crater on Ryugu, revealed that the cohesion of the subsurface layer on Ryugu is much smaller (140–670 Pa) than previously assumed, and that crater formation on Ryugu occurs in the gravity regime rather than in the strength regime that was assumed in previous studies (Arakawa et al., 2020). Because the surface cohesion alters the projectile-to-crater diameter ratio in crater production models, low cohesion conditions are required in these models to reproduce the observed crater distribution. The newly-determined low cohesion of asteroids has led to uncertainties regarding whether a seismic shaking model in a gravity regime can explain the depletion in the crater size-frequency distributions (CSFDs) of asteroids.

The resurfacing and crater obliteration processes also modify the properties of surface materials. One example is the cosmic-ray exposure (CRE) ages of surface materials. Any variation observed in the measured CRE ages depends on the vertical mixing rate of the surface materials. For example, if the surface of an asteroid has not been resurfaced at all, the CRE ages of the grains will simply describe the time over which the grain has remained at a depth of less than 1 m since the asteroid was born because galactic cosmic rays can only penetrate 1 m into the soil (Nagao et al., 2011). In contrast, if vertical mixing deeper than 1 m is active on the asteroid because of regolith convection or impact cratering, young CRE ages are obtained. This is because cratering excavates the surface of an asteroid, transporting fresh subsurface materials to the surface. Resurfacing processes in turn transport the materials at the surface to the floor of a crater, where the surface materials are buried. In this case, the resurfacing rate of the cratering and obliteration will control the CRE age. Comparison between the resurfacing rate and the CRE ages of returned samples provides information about the geological processes taking place on the surface of an asteroid, such as vertical movements or the outflow of surface materials. However, the resurfacing rate of Ryugu has not yet been determined based on crater data and crater production functions (CPFs).

Another example of how geochemical information is modified by resurfacing processes is the degree of space weathering which changes reflectance spectra of asteroid surfaces due to irradiation of cosmic ray and micrometeorite bombardment. The mixing of the material from the surface with that of the subsurface can change the reflectance spectra of the surface of an asteroid by exposing fresh materials from below the surface. The spectra of asteroid surfaces are altered by solar heating and space weathering (Hiroi et al., 1996; Lantz et al., 2017). Telescopic observations of asteroid families suggest that space weathering on asteroids may occur in two stages. Although space weathering reddens asteroid surfaces rapidly for the first $\sim 10^6$ yr, the reddening continues much more gradually throughout the following $\sim 10^9$ yr (Vernazza et al., 2009). Based on the spectral color analysis of S-type asteroid families, the degree of space weathering on ordinary-chondrite-like surfaces in the main belt probably proceeds to reach a 50% completion in $\sim 1$–10 Myr, and it takes longer time to reach a fully weathered surface (Brunetto et al., 2015). In contrast, such a two-stage space-weathering process has not been reproduced in laboratory experiments yet. Ion irradiation experiments carried out in laboratories suggest very fast ($10^3$-$10^4$ yr) space weathering (Lantz et al., 2017), which is faster than $\sim 10^6$ yr. However, the very slow space weathering in the second stage has not been observed in the ion irradiation experiments. Impact gardening was proposed to be a possible process for the slow space weathering on asteroids (Vernazza et al., 2009; Shestopalov et al., 2013). Since mixing fresh subsurface materials with weathered surface materials reduces the effective rate of space weathering, the spectral rejuvenation processes that occur on the surface of an asteroid could be the key to reconcile these inconsistent timescales. In a previous modeling study that investigated the process of rejuvenation by resurfacing, the timescale over which the topmost few tens of micrometers are resurfaced was estimated at approximately $5 \times 10^3$ yr on 20 km-sized asteroids (Shestopalov et al., 2013). The timescale of resurfacing is larger for smaller asteroids (Shestopalov et al., 2013); this timescale (~$5 \times 10^3$ yr) is much shorter than that inferred by the ion irradiation experiments. However, the rapid resurfacing rate that is assumed to result from impact cratering/ejecta blanketing has not been verified using actual geological data (e.g., craters, topography, ejecta) obtained during asteroid missions.

In this study, based on the distribution of craters on Ryugu that were observed by Hayabusa2, we propose a novel approach to investigate the mechanism by which resurfacing occurs on asteroids by deriving the relationship between the crater diameter ($D$) and the crater retention age ($t$). We show that $t$ is proportional to $D^a$ and then estimate the index $a$ to assess if seismic shaking accounts for the observed crater distribution, that is, $t \propto D^2$ (Richardson et al., 2005). We investigate the $t - D$ relationship between the craters on four asteroids: Ryugu, Itokawa, Eros, and Bennu, using CSFDs and CPFs and discuss the mechanism by which resurfacing occurs on the four asteroids. We compare the timescale of resurfacing with that of space weathering to investigate the decrease in the apparent space weathering rate on asteroid surfaces. Finally, we describe the relationship between resurfacing rates and the CRE ages of returned samples.

## 2 Crater Size-Frequency Distributions for Ryugu, Bennu, Eros and Itokawa

The CSFDs of four small asteroids, i.e., Eros, Itokawa, Ryugu, and Bennu, were used in this study (Chapman et al., 2002; Hirata et al., 2009; Sugita et al., 2019; Walsh et al., 2019). For Ryugu, craters larger than approximately 4 m in diameter were counted and classified into 4 confidence levels (Cho et al., 2021), using images that were obtained using the telescopic optical navigation camera (ONC-T) onboard Hayabusa2 in an observation campaign that took place on 1 August 2018 at an altitude of 5 km with an image resolution of approximately 0.5 m/pix and observations over an area lying between 50°S and 40°N. A total of 322 craters were found on Ryugu, with approximately 200 new craters identified in addition to those in Sugita et al. (2019). Comparison between the CSFDs of the previous study and this study shows no significant difference in the distribution of craters larger than ~10 m in diameter (Fig. 1). In this study, we used the CC1-2 crater candidates on Itokawa (Hirata et al., 2009), CL1-2 crater candidates on Ryugu (Cho et al., 2021), and the "distinct" crater candidates (Walsh et al., 2019) on Bennu. The CSFDs of the four small asteroids are plotted in an R-plot with $N^{-0.5}/A$ error bars, where $N$ is the number of craters and $A$ is the surface area over which the craters are counted (Fig. 2). The depletion in small craters was consistently observed on all four asteroids.

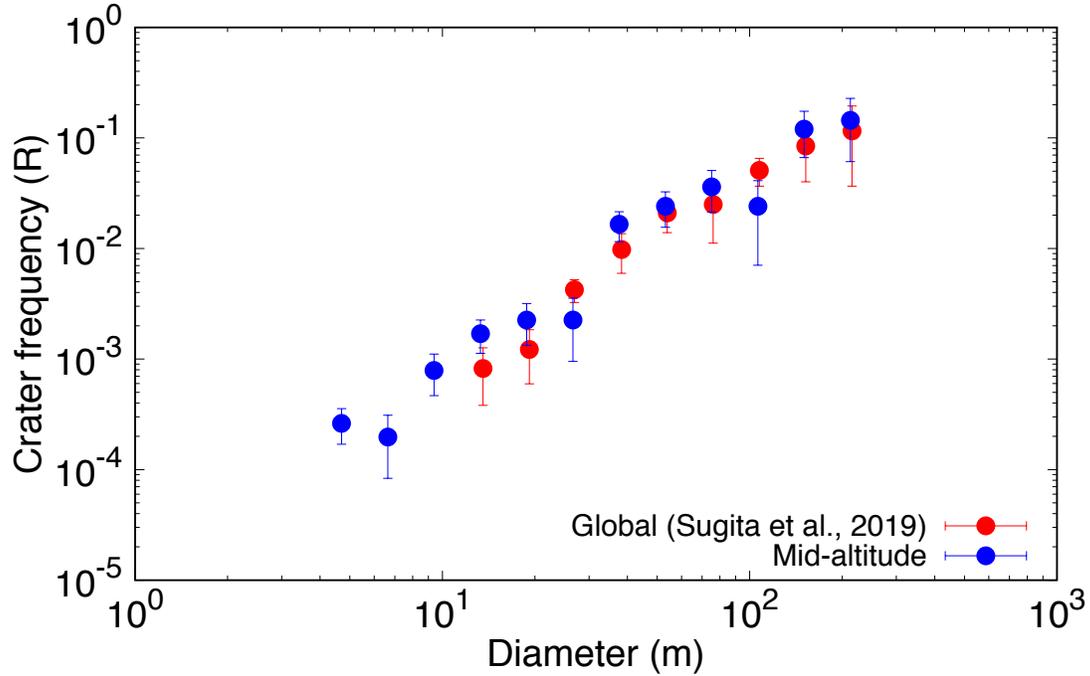

**Figure 1**. CSFDs of Ryugu obtained in this study and from a previous study in an R-plot. Red circles are craters counted by Sugita et al. (2019). Blue circles denote craters counted in this study. Our extended crater counting is consistent with the previous crater counting at $D > 10$ m.

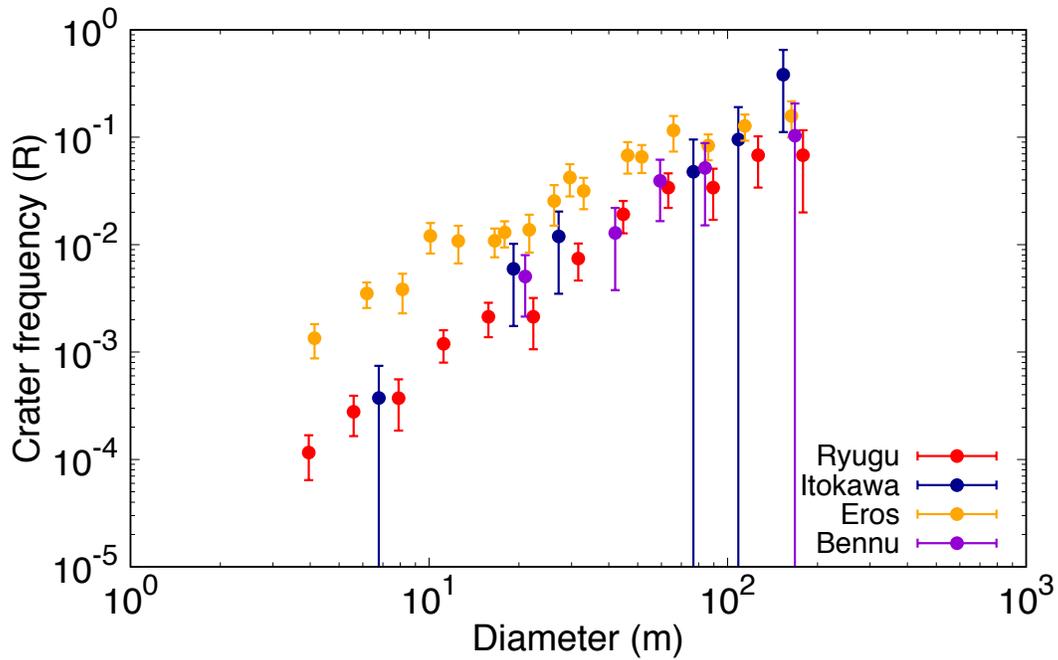

**Figure 2.** Crater size-frequency distributions (CSFDs; R-plot) on Ryugu, Itokawa, Eros, and Bennu. Error bars correspond to $N^{-0.5}/A$, where $N$ is the number of craters in a bin and $A$ is the surface area used in crater counting. Each bin is between $D/k$ and $kD$, where $D$ is the diameter at the center of the bin and $k$ is $2^{1/4}$.

## 3 Crater Retention Ages of Ryugu, Bennu, Eros, and Itokawa

In this section, we outline the crater production model and crater scaling relationship that were used for the analyses (Section 3.1). Second, we describe the method used for estimating crater retention ages (Section 3.2). We then show the relationship between crater diameter $D$ and crater retention age $t$. Finally, we discuss the power-law indices and the intercepts of the age-diameter relation (Section 3.3).

### 3.1 Crater production functions

CPFs reproduce the CSFDs of asteroids. To investigate the mechanism and efficiency of the resurfacing processes using CSFDs, the CPFs for each asteroid are first derived. The CPFs are formulated using impactor models and a crater scaling relation. In this study, two impactor models were used to address the uncertainties in the impactor models. Furthermore, we assume gravity scaling including the armoring effect (Tatsumi and Sugita 2018).

#### 3.1.1. Impactor size-frequency distribution models

Two impactor size distribution models were used to estimate the number of impacts per year $N_p(D_i)$. To estimate $N_p(D_i)$, four parameters, namely the diameter of the asteroid $D_a$, the diameter of the impactor $D_i$, the probability of collision $P_i$, and the number of potential impactors $N_i(D_i)$ in the main belt or near-Earth orbits, were used.

The probability of a collision occurring during a particular year per cross-section of an asteroid $P_i$ is estimated as $P_i = 2.86 \times 10^{-24}$ m$^{-2}$ yr$^{-1}$ for collisions with main belt asteroids (MBAs) and $15.3 \times 10^{-24}$ m$^{-2}$ yr$^{-1}$ for collisions with near-Earth asteroids (NEAs) (Bottke et al., 1994a). We assumed a single $P_i$ value for our calculation for the following reason. Because the mean intrinsic collision probability $P_i$ depends on the orbital elements of an asteroid, the use of a single $P_i$ model is not very good in the light of assessment of its uncertainty. However, the estimated uncertainty (a factor of <10) in collision probability $P_i$ (Bottke et al., 1994) is much smaller than that of main-belt impactor size frequency distribution models, as discussed in the following. Thus, the effect of the uncertainty in collision probability likely plays only a minor role and will not significantly change our results. The number of impacts $N_p(D_i)$ onto an asteroid of diameter $D_a$ can be estimated by multiplying the collision probability $P_i$ by the number of impactors $N_i(D_i)$ and the cross-sectional area $(D_a/2 + D_i/2)^2$,

$$N_p(D_i) = P_i N_i(D_i) \left(\frac{D_a}{2} + \frac{D_i}{2}\right)^2. \tag{1}$$

The cumulative number of main-belt asteroids (MBA) and near-Earth asteroids (NEA) larger than $D_i$, where $N_i(> D_i)$, was estimated based on collisional and dynamical simulations that were carried out previously by Bottke et al. (2005; BOT model, hereafter) and O'Brien & Greenberg (2005; OBG model, hereafter). In addition, we used the impactor distribution of NEAs based on a survey simulation that was carried out by Harris & D'Abramo (2015; HDA model, hereafter). The cumulative number of MBA impactors is estimated as

$$N_i(> D_i) = 1.32 \times 10^{13} D_i^{-2.53} \text{ (BOT model)},$$

$$N_i(> D_i) = 1.48 \times 10^{13} D_i^{-2.96} \text{ (OBG model)},$$

while the number of NEA impactors is estimated as

$$N_i(> D_i) = 1.67 \times 10^{10} D_i^{-2.53} \text{ (BOT model)},$$

$$N_i(> D_i) = 3.10 \times 10^{11} D_i^{-3.18} \text{ (OBG model)},$$

$$N_i(> D_i) = 9.84 \times 10^{9} D_i^{-2.35} \text{ (HDA model)}.$$

An impactor size $D_i$ in the range of 1–100 m is used in the BOT model and 0.1–100 m is used in the OBG model (Fig. 3). The number of impactors $N_i(D_{i,k})$ can be calculated using $N_i(> D_{i,k})$ and $N_i(> D_{i,k+1})$ ($D_{i,k+1} > D_{i,k}$), where $k$ is an index denoting a bin number. $N_i(D_i)$ is described using

$$N_i(D_i) = N_i(> D_{i,k}) - N_i(> D_{i,k+1}).$$

For our analyses, we assumed that the power-law relations of the impactor distributions can be extrapolated to sizes of small impactors down to 0.01–0.1 m in diameter.

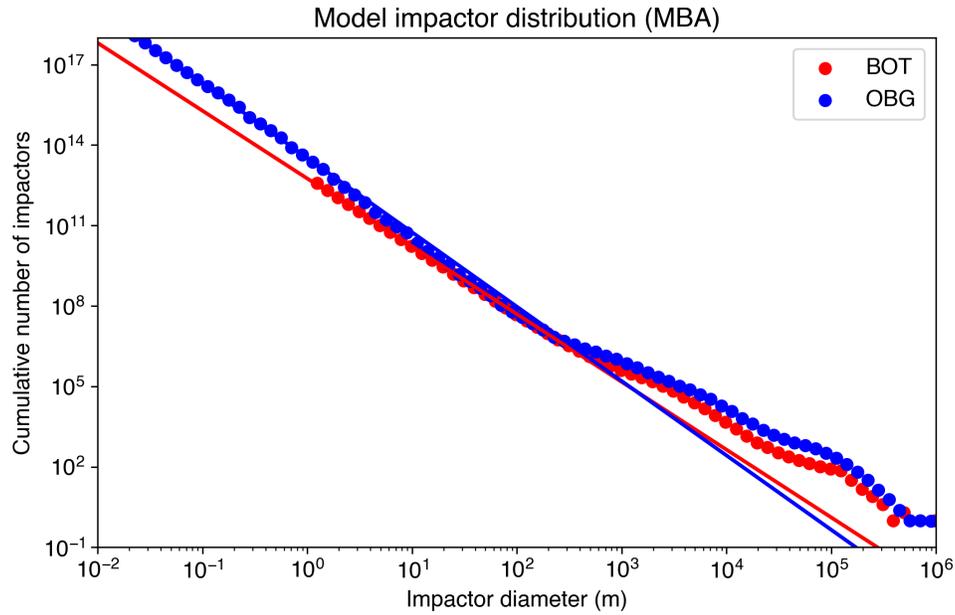

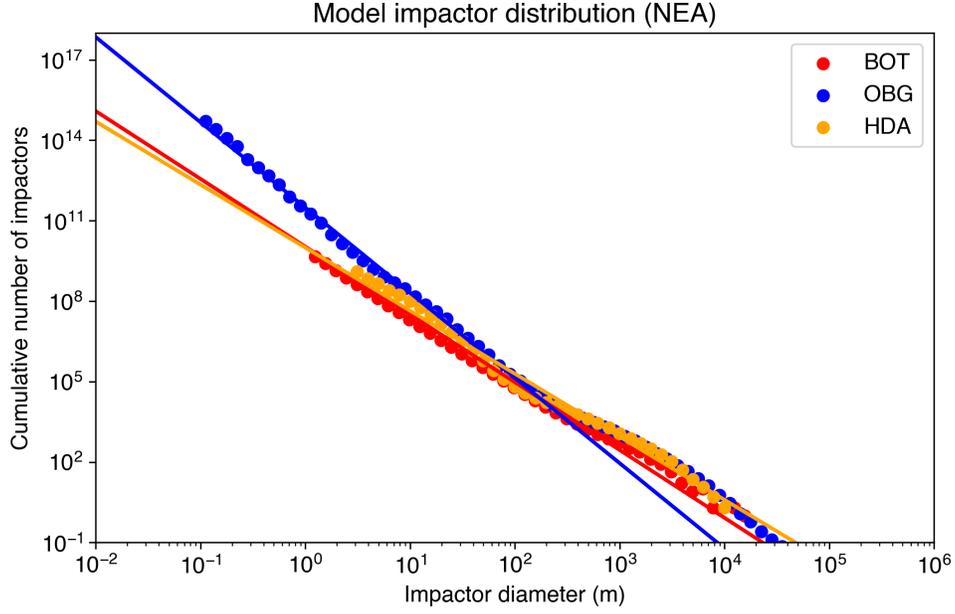

**Figure 3.** Distributions of MBA and NEA impactor model. The red, blue, and orange circles indicate the BOT, OBG, and HDA impactor models, respectively. The lines indicate fitting lines.

3.1.2. Crater scaling relation

The gravity crater scaling rule was used to estimate the diameter of a crater from the diameter of an impactor (Tatsumi & Sugita, 2018). The armoring effect was incorporated into the model to address the boulder-rich surfaces of Ryugu, Itokawa, and Bennu. The general crater scaling relation accounting for both strength and gravity as well as armoring effect is expressed as

$$\pi_V^* = \frac{\rho_t V_c}{m_p + m_t}$$

$$= K_1 \left[ \pi_2^* \pi_4^{-\frac{1}{3}} + K_2 \xi^{-\frac{(2+\mu_2)}{2}} + K_3 \pi_3^{*\frac{(2+\mu_3)}{2}} \right]^{-\frac{3\mu_1}{(2+\mu_1)}}, \qquad (2)$$

where

$$\pi_2^* = \Pi^{-\frac{7}{3}} \left( 3.22 \frac{r_p g}{U^2} \right),$$

$$\pi_3^* = \Pi^{-2} \frac{\bar{Y}}{\rho_t U^2},$$

$$\xi = \frac{m_p U^2/2}{m_t Q_D^*},$$

$$\pi_4 = \frac{\rho_t}{\rho_p},$$

$$\Pi = \frac{m_p}{m_p + m_t}.$$

where, $g$ is the surface gravity of the asteroid and $r_p$ and $U$ are the radius and velocity of the impactor, respectively. The median impactor velocity $U$ is 5.3 km/s in the main belt (Bottke et al., 1994a) and 18.5 km/s in near-Earth orbits (Bottke et al., 1994b). $\rho_p$ and $\rho_t$ are the densities of the impactor and the asteroid, respectively, while $m_p$ is the mass of the impactor. We assumed a disruption energy per unit mass, $Q_D^*$, of 1000 J/kg (Flynn, 2004) for a boulder that exhibits the armoring effect. $\bar{Y}$ is the effective strength, and most previous studies have used the dry-soil cohesion condition ($\bar{Y} = 0.18$ MPa, $\mu_3 = 0.24$ and $K_3 = 1$) for the crater scaling relation. However, the SCI experiment on Ryugu implied a low cohesion ($1.4 \times 10^{-4}$–$6.7 \times 10^{-4}$ MPa) of the surface in the impact area (Arakawa et al., 2020). Assuming this area is representative of the whole asteroid surface, we thus used a cohesionless condition ($K_3 = 0$) in our crater scaling relation. For angular-shaped grain targets, the scaling parameters $K_1$, $K_2$, $\mu_1$, and $\mu_2$ are 0.24, 0.01, 0.41, and 1.23, respectively (Tatsumi & Sugita, 2018). The mass of the target boulder $m_t$ determines the degree of armoring because $m_t$ includes a typical boulder size $\delta_t$ of $m_t = \frac{4}{3}\pi\delta_t^3\rho_g$, where $\rho_g$ is the boulder density. A boulder density of $\rho_g = 2300$ kg/m³ and an impactor density of $\rho_p = 2300$ kg/m³ (Sugita et al., 2019) were used. The parameters used to describe each asteroid are listed in Table 1.

To underline the importance of cohesion, the scaling relation under the dry-soil cohesion conditions was compared with that of cohesionless conditions (Fig. 4). An impactor of a given diameter will produce a crater approximately three times larger under the cohesionless conditions than it will under the dry-soil cohesion conditions. The crater size scaling relation under the cohesionless conditions is wavy due to the armoring effect (Fig. 4). The wavy feature of the scaling relation is controlled by the disruption energy per unit mass $Q_D^*$ (Tatsumi and Sugita 2018). A smaller disruption energy yields a less wavy scaling relation. Although a similar wavy feature is found with $Q_D^* = \sim 500$, the scaling relations are close to a linear line with $Q_D^* < 500$. However, the true value of $Q_D^*$ of asteroids' grains is not constrained yet by observations. We assumed that $Q_D^*$ is 1000 J/kg based on impact experiments for carbonaceous chondrites (Flynn, 2004).

**Table 1.** Properties of asteroids used to derive crater production functions (CPFs). $D_a$ is the diameter of the asteroid, $g$ is the surface gravity of the asteroid, $\rho_t$ is the density of the asteroid, and $\delta_t$ is the typical size of the boulders on the asteroid.

|  | Ryugu[a] | Itokawa[b] | Eros[c] | Bennu[d] |
|---|---|---|---|---|
| $D_a$ (m) | 896 | 326 | 16000 | 488 |
| $g$ (m s$^{-2}$) | $1.5 \times 10^{-4}$ | $1.0 \times 10^{-4}$ | $5.9 \times 10^{-3}$ | $6.15 \times 10^{-5}$ |
| $\rho_t$ (kg m$^{-3}$) | 1190 | 1900 | 2670 | 1190 |
| $\delta_t$ (m) | 3.0 | 2.0 | 0.0 | 2.0 |

[a]Watanabe et al., 2020; Sugita et al., 2019
[b]Fujiwara et al., 2006; Tatsumi & Sugita, 2018
[c]Robinson et al., 2002
[d]Barnouin et al., 2019; Walsh et al., 2019

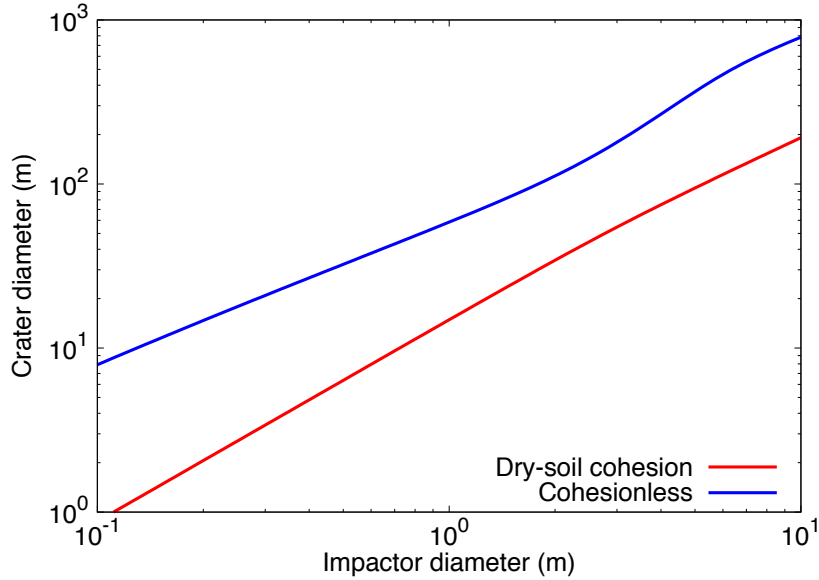

**Figure 4.** Relationship between the impactor diameter and crater diameter on Ryugu. The red line is calculated with dry-soil cohesion ($\bar{Y} = 0.18$ MPa). The blue line is calculated assuming cohesionless conditions. Note that the resulting crater diameters differ by a factor of ~3.

3.2 Estimation of main-belt asteroid model ages

Here, we estimate crater retention ages with the impactor distributions of MBA (MBA model ages) using the intersections between CPFs and CSFDs. CPFs are produced by cratering on the fresh surface of an asteroid (blue lines in Fig. 5(a)). If no crater obliteration is present, the crater distribution on an asteroid follows a single CPF. The time $t(D)$ required to produce the observed crater number density on an asteroid is obtained by calculating the intersections between the CPFs and CSFDs. Here, $t(D)$ is the retention age of craters of diameter $D$. Craters formed before $t(D)$ are considered obliterated and are therefore unobservable. We calculated the CPFs of different ages to estimate the effective retention ages for craters of different sizes on the CSFDs.

The MBA model ages of craters were derived as a function of diameter, indicating that the model ages of the craters on Ryugu increased with crater size instead of being represented by a single age (Fig. 5(b)). Our analysis further indicates that the relationship between the model age $t$ and the crater diameter $D$ ($t - D$ relationship) can be expressed as $t \propto D^a$ (Fig. 5(b)). The frequency of craters smaller than 200 m in diameter on the four asteroids decreased significantly as the crater size decreased, as did the model age of the craters.

Several reasons could potentially explain these depletions, such as a deficiency in small crater formations, observational bias for small craters, or the rapid erasure of smaller craters. We argue that the depletion in small craters is attributed to the erasure of craters via resurfacing processes, because the armoring effect, uncertainty of the impactor models, and observational bias for small craters are unlikely for craters that fall into the size range 4–200 m for the following reasons.

First, the various boulder sizes that contribute to the armoring effect may not explain the depletion in small craters. A typical grain size used in our model changes the degree of armoring effect: larger typical grain size decreases a resulting crater diameter. If the armoring effect alone can account for the depletion in the crater distribution, the CPF should follow the crater size frequency distribution. However, the CPFs do not follow the Ryugu's crater size frequency distribution if we assume a single boulder size of 3 m for the armoring effect (Fig. 5(a)). This result does not change even if we assume two boulder size regimes ($\geqq$ 3 m and < 3 m) which simulate an actual boulder size distribution. Here, we evaluate the number density of 10-m craters that would have been formed taking different boulder sizes into account. We take craters of 10 m on Ryugu as a typical size of depleted "small" craters. If we assume a single boulder size of 3 m for the armoring effect, the number of such small craters formed during 8.5 Myr (Morota et al. 2020) would have reached the empirical saturation line of crater frequency R~0.2 (Fig. 5a). We investigate whether the armoring effect can explain the degree of the depletion from the saturation level when different boulder sizes are adopted. Here we consider the two boulder size regimes ($\geqq$ 3 m and < 3 m), because the half of the Ryugu's surface is covered by the boulders $\geqq$ 3 m (Michikami et al. 2018; Sugita et al. 2019; Tanabe et al. 2021). In the former regime, 10-m craters cannot be formed since most of the impact energy is consumed by the destruction of boulders (Tatsumi and Sugita, 2018). As a result, the number density of the 10-m craters would have been half compared to the outcome of the single boulder size assumption (3 m). The other half of the surface of Ryugu is covered by the boulders of the second size regime (< 3 m). When an impact hits this surface, the number density of 10-m craters increases from that of the single boulder size assumption (3 m) because of a smaller armoring effect. Combining the effects of the both regimes, the number of 10-m craters can decrease by half at most. Thus, the armoring effect alone cannot explain the observed depletion in craters of more than three orders of magnitude.

Second, the impactor models are an unlikely factor. Our parameter study indicates that the power-law index of an impactor distribution needs to be shallower than -1.8 to account for the depletion in small craters in the R-plot; if the power-law index of an impactor distribution is -1.8, a CPF would be almost flat in an R-plot. However, the larger power is inconsistent with those in the BOT (-2.53 for both MBAs and NEAs) or OBG models (-2.96 for MBAs and -3.18 for NEAs). In addition, the distribution of NEAs is estimated based on other observations and analyses. The power-law index of the NEA impactor distribution in 10 m -10 km was estimated to be $N_i(>D_i) \propto D_i^{-2.35}$ based on survey simulations (Harris & D'Abramo, 2015). The cm-sized NEA impactor distribution was estimated to be $N_i(>D_i) \propto D_i^{-2.7}$ based on the observation of the flashes from meteoroid impacts on the Moon (Suggs et al., 2014). Because these power-law indices also differ from -1.8, the impactor size-frequency distribution cannot explain the depletion shown in the R-plot.

Third, crater observation bias would not account for the depletion. The completeness in a crater population can be lost for craters smaller than 10 pixels (Robbins et al., 2014), which corresponds to 5 m in this study. Nevertheless, the completeness in crater counting does not change this conclusion because the positive trend of the $t - D$ relationship (Fig. 5(b)) continues through all size bins, all of which were above 5 m in diameter. Thus, the $t - D$ relation (Fig. 5(b)) suggests that smaller craters are obliterated more rapidly by resurfacing processes than larger craters.

Estimating the retention ages of craters with different diameters provides information on the resurfacing mechanism from the power-law indices $a$ of the crater distribution. For example, if seismic shaking was at the origin of the depletion, the crater retention age (i.e., the time required to decrease the crater depth to 0.0005 $D$) should be proportional to the square of the crater diameter (i.e., $t \propto D^2$) (Richardson et al., 2005). This is because the crater degradation caused by this process is expressed as a diffusion process. Alternative resurfacing processes include surface flows induced by surface potential changes and tidal effects (Yu et al., 2014). The rate of resurfacing that results from surface flows may be controlled by the thickness and frequency of the flows, but the $t - D$ relation resulting from surface flows does not necessarily follow a power-law relation.

The power-law index $a$ of the $t \propto D^a$ was calculated by applying a bootstrap method (Efron, 1979) to the $t - D$ data points. First, we randomly sampled data points from the original data. The number of data points in a replicate dataset was equal to the number of original data points. A regression line and the corresponding power-law index were then derived from the replicated data set. The routine was repeated to derive the distribution of the best-fit power-law indices. The average and standard deviation of $a$ were then estimated.

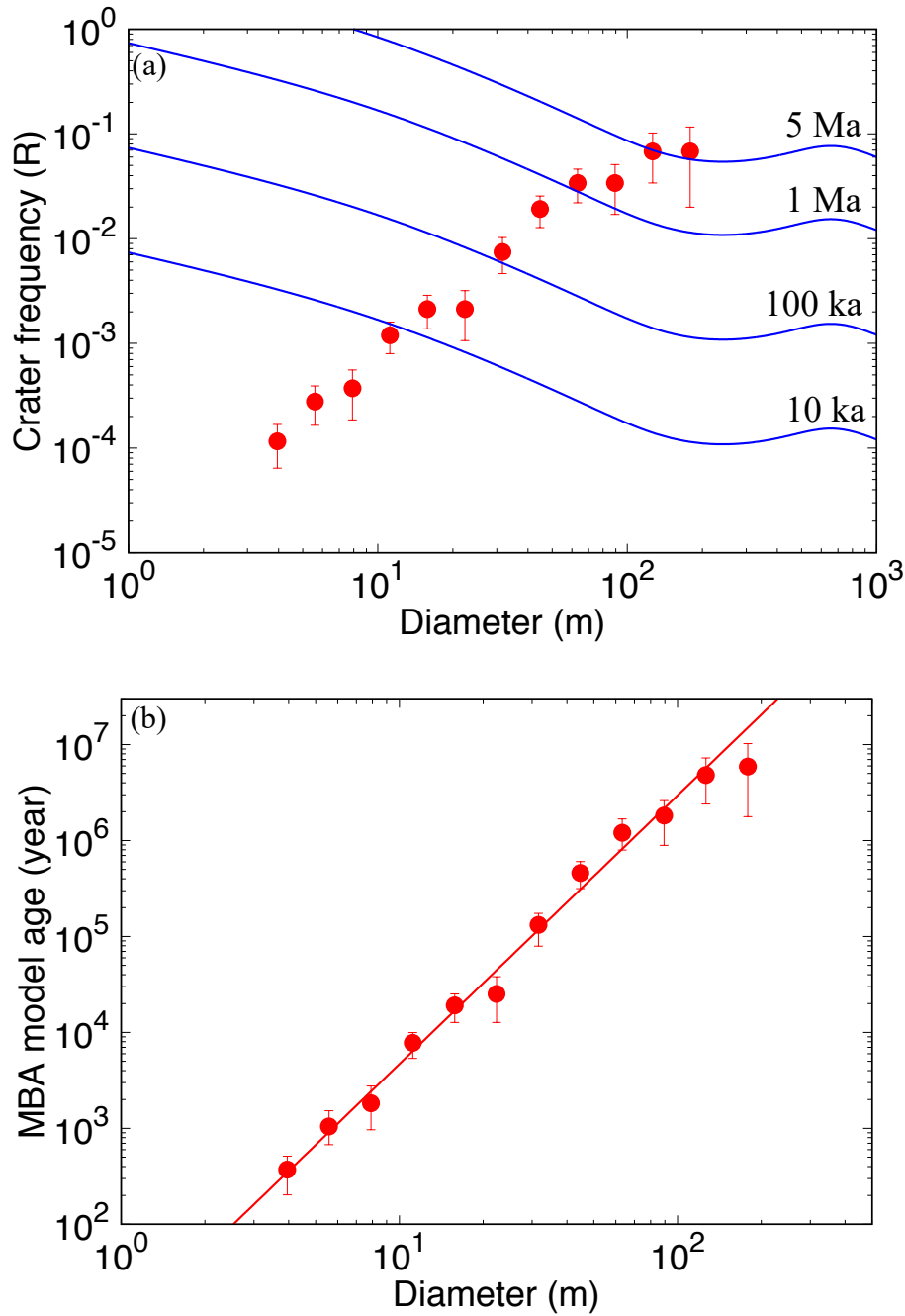

**Figure 5**. **(a)** CSFD of Ryugu (red circles) and CPFs of different ages (blue lines). **(b)** Relation between the MBA model age and crater diameter. The red line shows linear fitting of the log-log plot. It is obtained by calculating the intersections between CSFD and CPFs in plot (a).

### 3.3 Crater age-diameter relations

The MBA model ages of craters with different diameters were estimated for Ryugu, Itokawa, Eros, and Bennu. Although we assumed the main-belt asteroid populations as an impactor model, the 10-m craters on the asteroids may have been formed at near-Earth orbits because their retention ages are younger than the typical dynamical lifetime of NEAs (~10s Myr; Bottke et al. 2002). We will discuss the retention ages considering the transit from the main belt to near-Earth orbits in Section 4.2. Figure 6 shows the relationship between the MBA model age and the crater diameters for the four considered asteroids. If no crater obliteration is present, the relationship between the MBA age and crater diameter would be a straight horizontal line, which corresponds to an isochron in a CSFD. We compared the power-law indices of asteroids with that of a diffusion process ($a = 2$). Figure 6(a) is calculated assuming cohesionless conditions, while Fig. 6(b) is calculated assuming dry-soil cohesion ($\bar{Y} = 0.18$ MPa (Holsapple, 1993)), using the BOT model as the impactor distribution. Under the cohesionless condition, which is consistent with the result of the Hayabusa2 SCI experiment, the power-law indices $a$ are $2.7 \pm 0.1$ on Ryugu, $2.9 \pm 0.2$ on Itokawa, $2.5 \pm 0.4$ on Bennu, and $2.5 \pm 0.1$ on Eros. These power-law indices are significantly larger than those produced by a diffusion process in which $a = 2$. Furthermore, the power-law indices of the four asteroids coincide within the range of the fitting error. In contrast, under dry-soil cohesion conditions ($\bar{Y} = 0.18$ MPa), which were typically assumed before the Hayabusa2 impact experiment, the power-law indices $a$ are $2.0 \pm 0.1$ on Ryugu, $2.2 \pm 0.3$ on Itokawa, $1.8 \pm 0.3$ on Bennu, and $2.1 \pm 0.1$ on Eros. These power-law indices are consistent with those of the diffusion process.

The different scaling laws yield the different power-law indices. This is because the $D_{crater}/D_{impactor}$ is constant for strength-controlled cratering, whereas $D_{crater}/D_{impactor}$ varies with D_impactor for gravity-controlled cratering. The different ratios of the crater size to the impactor size lead to the different power-law indices in the $t \propto D^a$ relation. For example, the MBA model ages of 100 m-size craters under the strength-controlled surface condition are 10 times larger than those under the gravity-controlled surface condition; the MBA model ages of 10 m-size craters under the strength-controlled surface condition are 100 times larger than those under the gravity-controlled surface condition. Thus, the different assumptions on cohesion results in the different power-law indices.

Figure 6(c) shows the relationship between the MBA model age and the crater diameter using the OBG model as an impactor distribution model assuming cohesionless conditions. Using the OBG model increases the power-law indices (i.e., the slope of the lines in the plot) as compared to those generated using the BOT model (Table 2). Thus, we find that cohesionless surface material leads to power-law indices $a$ that are consistently larger than 2, regardless of which impactor size-frequency distribution model is employed.

The intercepts between the fitting lines in the age-diameter plots and the vertical line at $D = 10$ m indicate the retention ages of craters of 10 m in diameter. If we assume the BOT impactor population model and cohesionless crater scaling, the retention age of 10-m craters on Eros (~$3 \times 10^4$ yr) is older than on Ryugu and Bennu (~$3 \times 10^3$ yr) even when the uncertainty of crater counting is considered (Fig. 6(a)). If a future study reveals different impactor models and surface conditions of Eros, however, then the age value would be updated accordingly. The uncertainties in the MBA model ages due to impactor population models and crater scaling laws can be perceived by comparing Figs. 6(a)-(c). Although it is unclear why the retention ages of 10 m-size craters are longer on Eros, the size as well as the surface and internal structures of the asteroid may control the resurfacing rate, making it more difficult to erase craters of this size on Eros than on smaller asteroids like Bennu and Ryugu. Moreover, if equilibrium is not achieved (i.e., craters keep forming on a surface), the different retention ages of the asteroids simply reflect the different resurfacing events on the asteroids. The surface exposure age of Eros to the main belt was estimated to be $225 \pm 75$ million years using a "cuberoot" scaling law (Richardson et al., 2020) and 2.3-3.8 billion years using a Marchi scaling law (Bottke et al., 2020) based on the crater distribution on Eros, while the MBA model age of Ryugu was estimated to be ~10 Myr under the cohesionless condition by crater counting (Sugita et al., 2019). Therefore, the higher number density of craters on Eros than that on Ryugu may reflect the longer surface exposure age of Eros than that of Ryugu. Another possible explanation for the different retention ages is the difference in the orbital evolution of these asteroids. The 10-m craters on these asteroids may have been formed at near-Earth orbits because the retention ages of 10-m craters are young. In our model, we assumed that the impactor flux on these asteroids is the same. However, the impactor flux of NEAs could depend on their orbital evolutions. Eros does not currently cross the Earth's orbit (Michel et al., 1996), unlike Ryugu and Bennu (Wada et al., 2018; Lauretta et al., 2017), and these asteroids may have had different impact fluxes in the recent past.

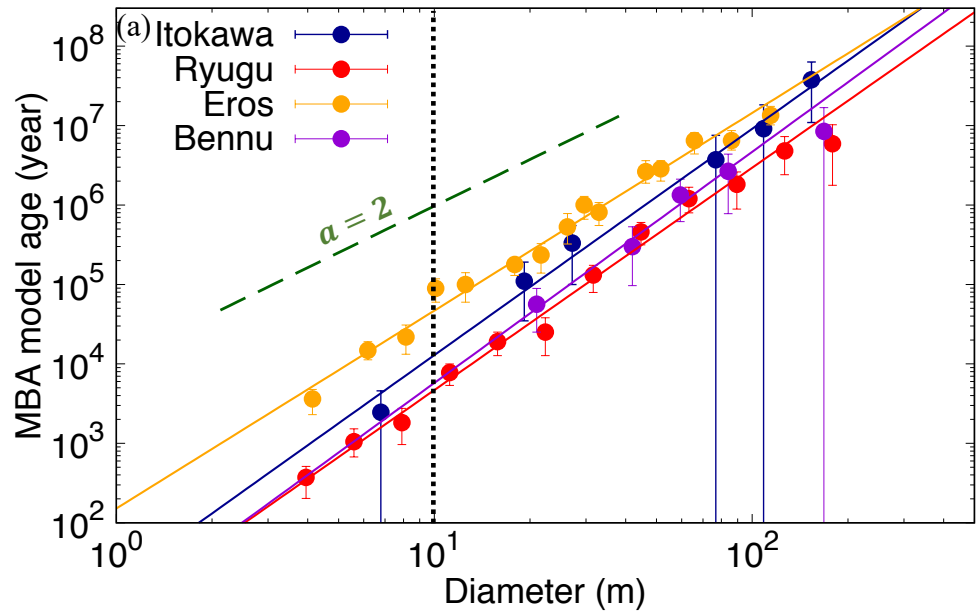
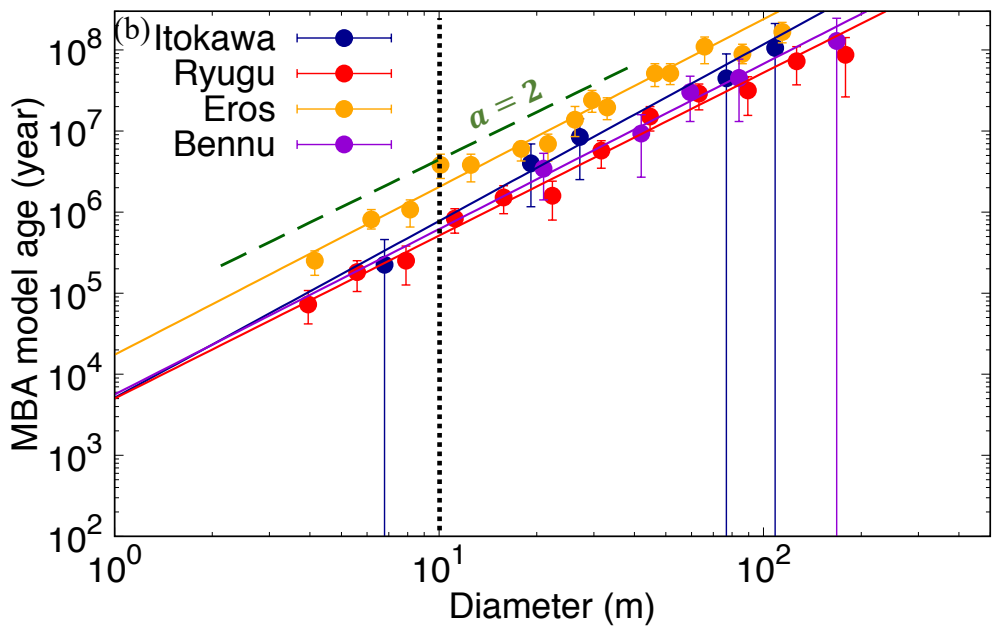

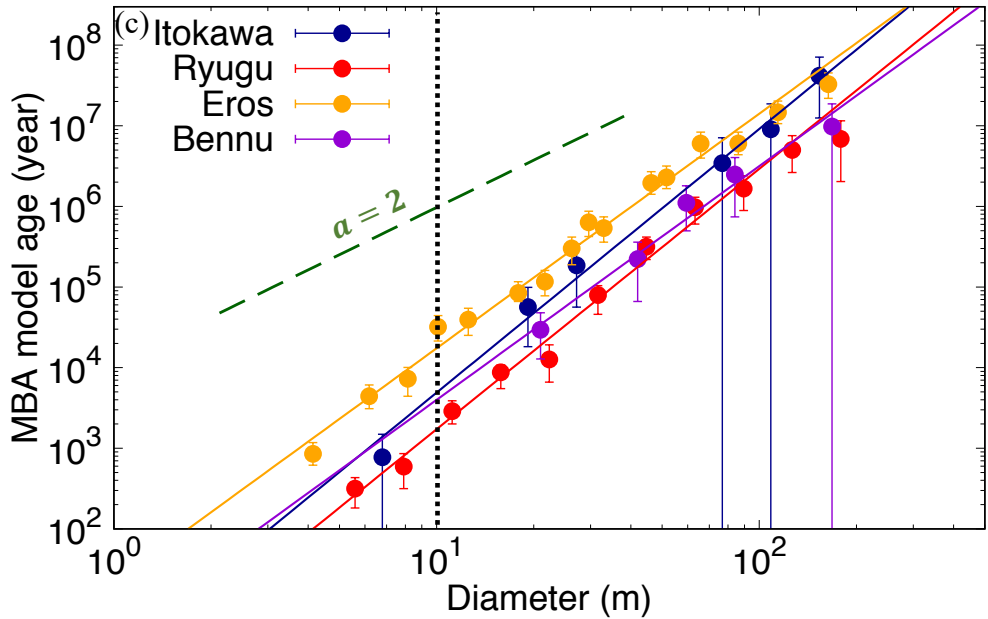

**Figure 6. (a)** Relation between main-belt asteroid (MBA) model age and crater diameter calculated using the BOT model assuming cohesionless conditions. The green broken line is overlaid to show the slope describing $a = 2$ (diffusion process). **(b)** Relation between MBA model age and crater diameter calculated using the BOT model and dry-soil cohesion conditions ($\bar{Y} = 0.18$ MPa). **(c)** Relation between MBA model age and crater diameter calculated using the OBG model and cohesionless conditions. Black dotted lines indicate D = 10 m.

**Table 2.** Power-law indices $a$ for the four asteroids calculated from the different impactor models (BOT model or OBG model) and assumed levels of cohesion ($\bar{Y} = 0$ MPa for cohesionless surface and $\bar{Y} = 0.18$ MPa for dry-soil cohesion).

|  | Ryugu | Itokawa | Bennu | Eros |
|---|---|---|---|---|
| $a$ (BOT model, cohesionless) | 2.7 ± 0.1 | 2.9 ± 0.2 | 2.5 ± 0.4 | 2.5 ± 0.1 |
| $a$ (OBG model, cohesionless) | 3.2 ± 0.1 | 3.3 ± 0.3 | 2.9 ± 0.4 | 2.9 ± 0.2 |
| $a$ (BOT model, dry-soil cohesion) | 2.0 ± 0.1 | 2.2 ± 0.3 | 1.8 ± 0.3 | 2.1 ± 0.1 |

## 4 Discussion

In this section, the implications of resurfacing processes and resurfacing rates on asteroids are discussed based on the results obtained from our analysis. More specifically, we discuss the possible reasons for the large power-law indices of the crater age-diameter relations in section 4.1, and the rate at which resurfacing occurs on Ryugu in section 4.2.

4.1 Large power-law indices of crater age-diameter relations

We discuss the power-law exponent using the cohesionless condition (Figs. 6(a) and (c)) because the SCI experiment showed that the cohesion of Ryugu's subsurface is much smaller (0.14 × $10^{-3}$–0.67 × $10^{-3}$ MPa) than previously assumed in a strength-controlled cratering (e.g., 0.18 MPa). The above analyses based on cohesionless conditions (Figs. 6(a) and (c)) indicate that the power-law exponent for the $t - D$ relations is significantly greater than 2, which is inconsistent with the prediction from the standard seismic shaking models. However, this does not necessarily mean that seismic shaking did not play an important role in crater erasure on Ryugu. In the following, we discuss three possible reasons, including modification of seismic shaking models, which may cause the power-law indices to be greater than 2: different CPF models, different diffusion coefficients in the seismic shaking models, and resurfacing processes that differ from the diffusion process, such as mass movement.

The first possibility is that the CPF models assumed in this study differ from the actual crater production. The power-law indices of the relations between the MBA model age and crater diameter can vary depending on parameters that are included in the CPF model, such as the impactor size-frequency distribution, armoring effect, and the cohesion conditions (Table 2). Note that the degree of the armoring effect was estimated to be insufficient to account for the depletion in small craters in Section 3.2. Then, a shallower slope of the impactor size-frequency distribution than that assumed in our model can yield small power-law indices. In this study, due to lack of data at small impactor sizes, we extrapolated the slope describing the distribution of impactors that are larger than a few meters in diameter to the cm-size impactor distribution. Our results suggest that the slope of the small impactor distribution should be shallower (i.e., $N_i(> D_i) \propto D_i^{-2.1}$) than that generated by the impactor distribution models ($N_i(> D_i) \propto D_i^{-2.53}$ and $N_i(> D_i) \propto D_i^{-2.96}$) to yield power-law indices that are close to those of the seismic shaking model (Fig. 7). In this hypothetical model, the power-law index of Itokawa (2.4±0.2) is a little larger than 2, but the other power-law indices (2.1±0.1 for Ryugu, 1.9±0.4 for Bennu, and 1.9±0.1 for Eros) are consistent with those of the seismic shaking model. Even if the cohesionless condition is used in our models, the impactor model with a shallower size-frequency distribution slope yielded the power-law indices close to 2. However, the cm-sized NEA impactor distribution is estimated to be $N_i(> D_i) \propto D_i^{-2.7}$ based on the observation of the flashes on the Moon (Suggs et al., 2014). Therefore, an actual cm-sized NEA distribution is less likely to be proportional to $D_i^{-2.1}$.

Furthermore, assuming a dry-soil cohesional surface (0.18 MPa), the power-law indices are approximately 2 and are therefore consistent with the seismic shaking model (Table 2). However, the cohesive strength of the subsurface layer of Ryugu was estimated to be between 1.4 × $10^{-4}$ MPa and 6.7 × $10^{-4}$ MPa from the SCI experiment (Arakawa et al., 2020). Therefore, assuming that the result of this experiment is representative of the whole Ryugu's surface, using a surface with dry-soil cohesion (0.18 MPa) is not suitable for estimating the crater retention ages on Ryugu. We assumed that the cohesion on Bennu would be similar to that of Ryugu because of the similar morphology (e.g., raised rim and bowl shape cavity of craters and a boulder-rich surface; Walsh et al., 2019). An evaluation of the cohesion on Eros and Itokawa is difficult because the properties of them, such as composition, size, and density, are different from those of Ryugu. If actual cohesion conditions on Bennu, Itokawa and Eros are similar to the dry-soil cohesion condition, the power-law indices are consistent with the seismic shaking models ($a = 2$).

The second possibility is that the actual crater obliteration process is not expressed as a simple linear diffusion equation, in which a constant diffusion coefficient is used under the assumption that the surface regolith flux is proportional to the topographic slope. For example, if the mechanical strength or the particle size at the surface of an asteroid is different from that at subsurface, the diffusion coefficient of the subsurface layer could also be different. If the diffusion coefficient decreases with increasing depth, materials located at greater depths are more difficult to move than those on the surface. In this case, materials near the floor of large craters have low mobility; thus, larger craters will have longer retention ages. A small pit found in the center of the SCI crater suggests that the subsurface layer of Ryugu has stronger cohesion than the surface layer (Arakawa et al., 2020). The longer retention ages of large craters increase the power-law index by rendering the plots steeper, as shown in Fig. 4. Detailed crater shape analyses are required to investigate the vertical profile of the diffusion coefficient.

If the surface of asteroids is modified by movement of the regolith that results from seismic shaking, the intercepts in Fig. 6 may reflect the efficiency of seismic shaking on each asteroid. The difference in the intersections suggests that the efficiency of seismic shaking on Ryugu and Bennu is greater than that on Eros (Fig. 6). On smaller asteroids, impacts are assumed to easily cause acceleration greater than the gravity at the surface (Richardson et al., 2005). Our results support the idea that the surface gravity of asteroids controls the efficiency of seismic shaking. Observations of other larger planetary bodies, such as the Moon and Ceres, also agree with gravity effect. For example, CSFDs from 10 km to 100 km on lunar heavily cratered highlands and Ceres are similar to each other, and the CSFDs are close to the empirical saturation line (Strom et al., 2018). On the other hand, the distributions of small craters on small asteroids may be not saturated because R-values of small craters are lower than 0.1 (Fig. 1), which is the lowest limit of the empirical saturation (Richardson, 2009). The surface gravities of the Moon and Ceres are 100-1000 times larger than those of the small asteroids, such as Ryugu and Bennu. Thus, CSFDs on the Moon and Ceres suggest that the efficiency of seismic shaking may be lower on planetary bodies with greater surface gravity.

The third possibility is that resurfacing processes on small asteroids are not the result of diffusion processes. Geomorphological evidence of surface flows was found along the current geopotential gradient on Ryugu, from the equator towards the mid-latitudes (Sugita et al., 2019). Surface flow features were also found on Itokawa and Bennu (Miyamoto et al., 2007; Jawin et al., 2020). Such flows could have obliterated the craters on these asteroids. The rate at which craters are obliterated as a result of surface flows depends on the thickness and frequency of the flows. The $t - D$ relation of Ryugu's craters (Fig. 6) indicates that flows with a thickness of 2–4 m would have occurred once in $10^4$–$10^6$ yr on average for the erasure of craters because the MBA model ages of 20-m craters are $10^4$–$10^6$ yr (Fig. 6). Figure 6 further indicates that csraters smaller than 10 m have much shorter retention ages ($10^2$–$10^5$ yr). These model ages suggest that surface flows of tens of centimeters occur at least once in every $10^2$–$10^5$ yr.

The driving force of the surface flow includes changes in the surface geopotential gradient. The past spin rate of Ryugu is assumed to have been faster than its current rate (Watanabe et al., 2019). The change in the spin rate of Ryugu induces various surface geopotential gradients. Bennu also undergoes a change in spin rate due to the Yarkovsky-O'Keefe-Radzievskii-Paddack (YORP) effect (Hergenrother et al., 2019). If surface slopes on an asteroid exceed the angle of repose during the change in spin rate, surface materials on the slopes would move along the downslopes. However, if surface slopes do not exceed the angle of repose, some additional mechanism is probably required to cause surface flows, such as seismic shaking. The frequency of such surface flows is difficult to estimate because the frequency depends on the change in the spin rate and the surface potential conditions. Nevertheless, similar surface flows may occur on other asteroids that have undergone changes in the spin rate.

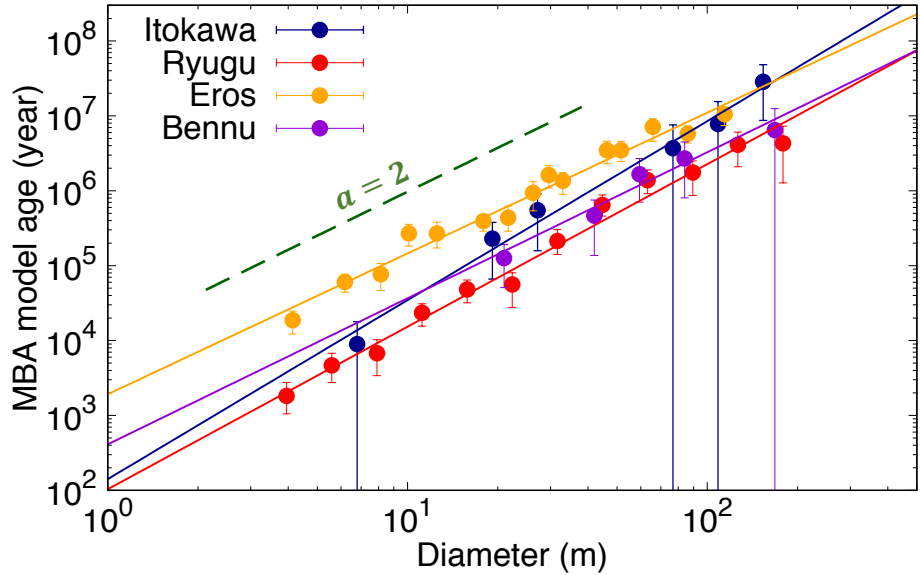

**Figure 7.** Relation between main-belt asteroid (MBA) model age and crater diameter calculated using the impactor model of $N_i(> D_i) \propto D_i^{-2.1}$ assuming cohesionless conditions. The green broken line is overlaid to show the slope describing $a = 2$ (diffusion process).

4.2 Relation between depth and retention age on Ryugu

In this section, we discuss the implications of the estimated crater retention ages on Ryugu in terms of returned sample analysis. A comparison between the expected CRE ages of returned samples and the crater retention ages can be used to interpret the obtained CRE ages. The obliteration of a crater $d$ m in depth requires material to migrate vertically over the entire depth of the crater: excavation by cratering transports subsurface materials to the surface of an asteroid, while crater obliteration leads to the transport of surface materials to the crater floors. If a crater is erased by seismic shaking, the regolith present on a crater wall slides down to the floor of the crater (Richardson et al., 2005). If a crater is erased by surface flow, surface materials outside the crater fill the crater. Surface materials move macroscopically from the top to the bottom regardless of the specific obliteration process, and the vertical scale of regolith migration is equal to the depth of the crater. The timescale of crater obliteration (lifetime of a crater) can be estimated as the retention age, whether the erasure process is seismic shaking or mass movement. Thus, the erasure timescale of a crater of a particular depth is equal to the residence time of the grains at the depth of the crater.

For a more accurate comparison between CRE ages and surface residence time, we show the retention ages of craters on Ryugu considering the transition from the main belt to a near-Earth orbit. When Ryugu migrated from the main belt to a near-Earth orbit, its orbit may have had a large eccentricity crossing the main belt. Then, most collisions will come from projectiles in the main belt; collisionally coupled to the main belt (Bottke et al., 1996). In contrast, when the aphelion of Ryugu becomes less than ~1.6–1.7 AU, most of the impacts on Ryugu would be by NEAs; collisionally decoupled from the main belt. Thus, the crater production rate greatly differs depending on whether Ryugu's orbit is coupled or decoupled from the main belt. We therefore calculated the relation between a crater retention age and crater depth for two different impactor population cases: all craters on Ryugu were formed by impacts of MBAs, and small craters on Ryugu were formed by impacts of NEAs.

Because Ryugu is estimated to have undergone its orbital evolution events, such as collisional decoupling from the main belt and orbital excursion toward the Sun (Morota et al., 2020), an appropriate crater production function should be used for each evolutionary stage. First, Ryugu is estimated to have come from the main belt (e.g., Michel and Delbo, 2010). When Ryugu is still in the main belt, all the craters on Ryugu were formed by impacts of MBAs. The retention age of craters on Ryugu are equal to MBA model ages, which were calculated in Section 3.3:

$$\text{Retention age} = \text{MBA (BOT) model age} \tag{3}$$

$$\text{Retention age} = \text{MBA (OBG) model age.} \tag{4}$$

Second, Ryugu is estimated to have an orbital transition from the main belt to near-Earth orbit most likely via $v_6$ resonance (e.g., Michel and Delbo, 2010). If Ryugu stays collisionally coupled to MBA after the cessation of the reddening event, the crater retention age of blue craters, which dominate craters < 30 m in diameter (Morota et al., 2020), can be estimated with the same equations (3) (4).

Third, if Ryugu collisionally decoupled from the MBAs before the cessation of the reddening event on Ryugu, then the retention ages of the blue craters can be estimated using the distribution of NEAs (NEA model ages). In fact, the period over which Ryugu has been in a near-Earth orbit was estimated using the BOT model ($8.1 \times 10^6$ yr) and OBG model ($0.3 \times 10^6$ yr) (Morota et al., 2020). The CSFD analysis using the craters smaller than 20 m also suggests the decoupling of Ryugu from the main belt $7.2 \pm 1.6$ Myr ago (Cho et al., 2021). Our study also shows that the retention ages of craters < 30 m on Ryugu are younger than the estimated period (Fig. 8), suggesting that the craters < 30 m on Ryugu were formed in the near-Earth orbit. Most of the interiors of craters smaller than 30 m (~70 %) have bluer spectra than their surroundings (Cho et al., 2021). The blue craters on Ryugu are considered to have been formed at the near-Earth orbit based on Ryugu's stratigraphy of surface colors (Morota et al., 2020). The other ~30% of the craters smaller than 30 m may have been formed in the main belt, which would decrease the estimated ages in Fig.9(b) by 30%. Nevertheless, this uncertainty does not change the resurfacing timescale of $10^3$–$10^5$ yr. The retention ages of craters < 30 m are expressed as NEA model ages depending on each impactor population model:

$$\text{Retention age} = \text{NEA (BOT) model age} \tag{5}$$

$$\text{Retention age} = \text{NEA (OBG) model age} \tag{6}$$

$$\text{Retention age} = \text{NEA (HDA) model age.} \tag{7}$$

Fourth, larger craters (e.g., > 44 m) are too large to be formed after Ryugu was decoupled from the MBAs even for the early decoupling case. Because a crater production rate on NEAs collisionally decoupled from the MBAs can be much lower than that for collisionally coupled NEAs, the retention ages for large (>44 m) craters should be expressed as the sum of the NEA retention age ($8.1 \times 10^6$ yr for the BOT model or $0.3 \times 10^6$ yr for the OBG model) and MBA retention age.

$$\text{Retention age} = \text{MBA (BOT) model age} + 8.1 \times 10^6 \text{ yr} \tag{8}$$

$$\text{Retention age} = \text{MBA (OBG) model age} + 0.3 \times 10^6 \text{ yr.} \tag{9}$$

Here, it is noted that the effect of different decoupling timing is minimal for the crater production model by O'Brien and Greenberg (2005) because the retention ages for the MBA model (equation (4)) and NEA model (equation (6)) based on O'Brien and Greenberg (2005) are very similar to each other.

Figure 9 shows the relationship between retention age and initial crater depth, which was derived from the crater diameter under the assumption that $d/D = 0.1$–$0.2$. Note that the NEA model ages of the HDA impactor model were calculated only for craters smaller than 30 m because an MBA impactor distribution is not estimated based on the same method as the HDA impactor model. The resurfacing timescale for craters that are shallower than 1 m is particularly rapid ($10^3$–$10^5$ yr) because of the frequent bombardment with impactors that can excavate craters of less than 1 m. Larger-scale impacts that would mix the surface with the subsurface are less frequent than the small-scale impacts, and the time required to resurface craters of a certain depth increases as the depth increases from $10^2$–$10^4$ yr (0.5 m) to $10^5$–$10^6$ yr (4 m) (Fig. 9). These results suggest that materials on the 1-m surface layer on Ryugu are unlikely to stay on the same layer for > 1 Myr.

The retention ages of the craters on Ryugu may provide clues to the observed young CRE ages of CM and CI chondrites. Ryugu surfaces exhibit spectra similar to those of moderately dehydrated CM and CI chondrites (Sugita et al., 2019). Noble gas analysis has indicated that the CRE ages of CM and CI chondrites are 0.1–1 Myr, which are significantly younger than the age of other carbonaceous chondrites or ordinary chondrites of 1–100 Myr (Eugster et al., 2006). If small (i.e., ≲1 m) fragments are ejected from the parent bodies in the main belt and migrate to collide with Earth as meteorites, the CRE age of such objects will be controlled by the collisional lifetime of meter-size objects, which is estimated to be ~7 Myr (e.g., O'Brien and Greenberg, 2005). However, this inference is based on the assumption that most meteorites come directly from the main belt, which may not be true particularly for carbonaceous chondrites. Although most meteorites have come from the main belt because their CRE ages (>8–10 Myr) are longer than the orbital lifetime of m-sized object in the near-Earth object (Eugster et al., 2006), CM and CI chondrites with short CRE ages (<7 Myr) may have come from objects in Earth-crossing orbits. In fact, trajectory analyses based on fireball observations for "fall" meteorites suggest that carbonaceous chondrites come from near-Earth orbits because most of their orbital migration timescale from the main asteroid belt are much longer than their CRE ages (Granvik and Brown, 2018).

If small fragments are ejected from the parent bodies in its near-Earth orbit, the collisional lifetime of the m-size objects in near-Earth orbits is likely to be much longer than 7 Myr because the rate of catastrophic collisions between near-Earth asteroids is 10–100 times smaller than that of collisions with main-belt asteroids (Bottke et al., 1993). The long collisional lifetime of meter-size NEAs (70-700 Myr) is much longer than the CRE ages of CM and CI chondrites (<7 Myr). In this case, the collisional lifetime does not set an upper limit to the CRE ages of CM and CI chondrites. If a collision does not reset the CRE ages, the observed CRE ages of m-size NEAs are the sum of its residence time on the parent body surface (≲1 m in depth) and migration time from the parent body to the Earth.

In fact, our results suggest that the average retention age of the top 1 m, which is recorded as a CRE age, is $10^3$–$10^5$ yr (Fig. 9). The upper estimate of the surface retention age (0.001–0.1 Myr) is comparable to the CRE ages of the least exposed CM/CI chondrites (i.e., 0.1–10 Myr). Because collisions between the Earth and a near-Earth body would have to occur before the orbital lifetime of the meteoroid (10s Myr; Bottke et al. 2002) ends, the CRE time accumulated during orbital migration after getting ejected from the near-Earth parent body should be ≲10s Myr. The residence time on the surface of parent bodies could be a primary contributor to CRE ages if they are as young as ~0.1 Myr. Note that NEAs with semimajor axes < 2 AU typically have long orbital lifetimes (10s Myr), but CM/CI chondrites with such long CRE ages have not been found yet (Eugster et al., 2006). Three possible reasons were discussed by Scherer and Schultz (2000) to account for their short CRE ages: (1) meteoroids migrated from the main belt to near-Earth orbits within a short period because their parent bodies were close to orbital resonance, (2) their parent bodies were in the Earth-crossing orbit when collisions ejected the meteoroids, and (3) their fragile character diminishes the rate of survival in space.

Based on our model, surface materials on parent bodies should be exposed to "$2\pi$" cosmic-ray irradiation. The "$2\pi$" exposure geometry of meteorites during the residence in a parent-body regolith may be detectable as neutron-capture effects and/or complex (multi-stage) exposure histories (Eugster et al., 2006). In fact, the evidence for the "$2\pi$" exposure geometry was reported for some enstatite chondrites based on the isotopic deviations of $^{149}$Sm-$^{150}$Sm and $^{157}$Gd-$^{158}$Gd (Hidaka et al., 1999). For the carbonaceous chondrite Murray (CM2), the isotopic ratios of $^{149}$Sm-$^{150}$Sm and $^{157}$Gd-$^{158}$Gd were measured, but no excesses of the isotopic ratios were reported (Murthy & Schmitt, 1963). However, the isotopic ratios of $^{149}$Sm-$^{150}$Sm and $^{157}$Gd-$^{158}$Gd have not been systematically investigated for other CM and CI chondrites. Since the samples returned from Ryugu and Bennu must have been exposed to the "$2\pi$" cosmic-ray irradiation, investigating Ryugu/Bennu samples would provide the insight for the proxy of "$2\pi$" exposure geometry of carbonaceous materials.

Our results for the young retention age ($10^3$–$10^5$ yr) on the surface of Ryugu help to resolve the inconsistency between the rapid space weathering process based on experiment ($10^3$–$10^4$ yr) (Lantz et al., 2017) and the slow space weathering based on observation ($10^6$–$10^7$ yr <) (Brunetto et al., 2015). The timescale for resurfacing that reaches depths shallower than 1 m on Ryugu is estimated to be $10^3$–$10^5$ yr, depending on the impactor model used (Fig. 9). The resurfacing timescale ($10^3$–$10^5$ yr) suggested by our results is comparable to or longer than the timescale of space weathering ($10^3$–$10^4$ yr) estimated by ion irradiation experiments (Lantz et al., 2017). The result of this comparison suggests that surface materials are mixed with subsurface materials, exhibiting a low effective space weathering rate. In addition, the timescale of space weathering on Bennu was estimated to be ~$10^5$ yr based on the analysis of crater colors (DellaGiustina et al., 2020). This timescale is longer than the rapid space weathering ($10^3$–$10^4$ yr) obtained in ion radiation experiments. The different timescales suggest that space weathering rate on Bennu is also decreased by resurfacing.

Our results suggest that the surface materials on Ryugu are subject to space weathering only for $10^3$–$10^5$ yr. Although this timescale is much shorter than that for spectral evolution of S-type asteroids (e.g., Vernazza et al., 2009), space weathering likely causes reflectance contrast between surface and subsurface on Ryugu even in the short timescale. In fact, the ejecta from the artificial crater was clearly darker than the surrounding surface, suggesting the presence of space weathering on Ryugu (Arakawa et al., 2020). When resurfacing processes mix surface and subsurface materials, the new surface materials are exposed to space and begin to experience space weathering. Through the repetition of this sequence, materials that experienced the weak weathering are transported to a few meters below the surface. This picture of slow space weathering of the entire shallow subsurface is consistent with the results of telescopic observations of asteroids that suggests slow space weathering (Brunetto et al., 2015). Our analysis based on the crater observation of Ryugu strongly suggests that slow space weathering due to the fast resurfacing by cratering, a process simulated using a theoretical model describing the generation and movement of regolith (Shestopalov et al., 2013). If such a surface gardening process occurs on Ryugu, it would lead to a gradual stratigraphy from more space-weathered surface to fresher subsurface. However, no such a gradual color stratigraphy has been found on Ryugu. The most outstanding red-blue stratigraphy found on Ryugu (Sugita et al. 2019) is unlikely caused by the combination between gardening and continual space weathering due to ion irradiation and/or meteoritic bombardment because the color stratigraphy has a rather sharp discontinuity, which results in the bimodal distribution of crater floor color. This stratigraphical color structure is likely to be caused by solar heating, which is a short period reddening event during an orbital excursion toward the Sun (Morota et al., 2020).

Comparison between the CRE exposure age distribution and the crater retention ages provides information on both the surface activity of Ryugu and the impactor distribution that formed the small craters (< 30 m). Our crater counts and impact flux model suggest that the mixing of the material at a depth of 2–4 m via impact cratering occurs once every $10^4$–$10^6$ yr on average (Fig. 9). Note that comparing the CRE ages with the resurfacing timescale of the cosmic-ray penetration depth (~1 m) is inappropriate because the mixing of material (boulders and pebbles) within the top 1 m of the subsurface allows the CRE ages to accumulate. Rather, we should compare the CRE ages of returned samples with the timescale required to resurface to a depth of 2–4 m because mixing at a depth of 2–4 m excavates fresh subsurface material and yields young CRE ages.

Taking the above caveats into account, the implications of the different CRE ages that may be measured from returned samples can be summarized. If the CRE ages of the Ryugu samples are measured to be 1–10 Myr (i.e., the sample was at a depth shallower than 1 m for 1–10 Myr in total), the retention ages using the HDA's NEA model would be comparable to the CRE ages, and the retention ages using the other NEA models would be underestimated by an order of magnitude. This underestimation will be attributed to the uncertainties of parameters used in our model, such as the single collisional probability, impactor velocity distribution, and crater scaling laws rather than that of the NEA impactor models. In fact, the NEA impactor distribution is well-constrained by observations (Brown et al., 2002; Suggs et al., 2014).

If the CRE age of the Ryugu samples is 0.1 Myr, this is consistent with our depth-age relations of Ryugu. In BOT's NEA model, the vertical mixing of materials that originate 2–4 m from the subsurface occurs once in approximately every $10^6$ yr on average (Fig. 9). This calculated timescale suggests that the surface materials remain within the top 1 m for < 1 Myr. In this case, the small craters on Ryugu would have been formed by an impactor population similar to that generated by the BOT model.

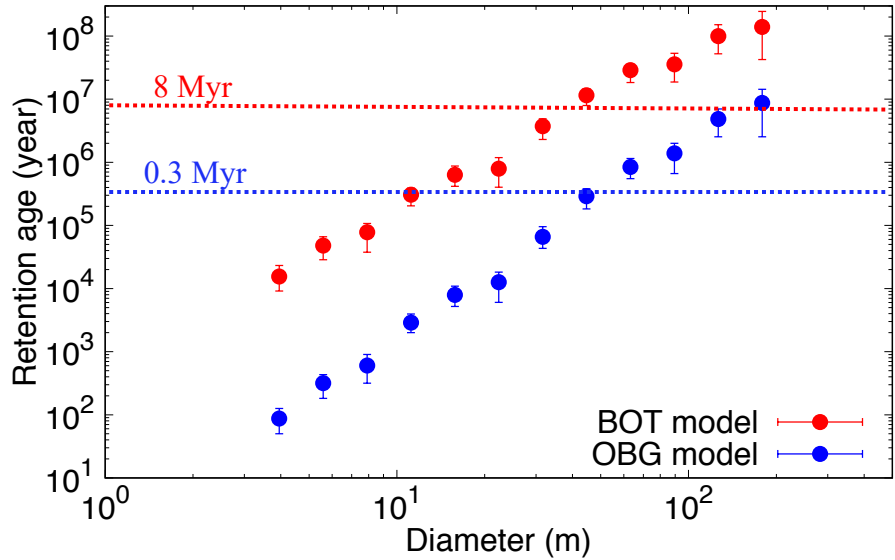

**Figure 8.** Relation between crater retention age and crater diameter on Ryugu calculated from different near-Earth asteroid distribution models (red: BOT model, blue: OBG model). The red dashed line (8 Myr) is the estimated time that Ryugu remained in a near-Earth orbit derived with the BOT model. The blue dashed line (0.3 Myr) is derived with the OBG model.

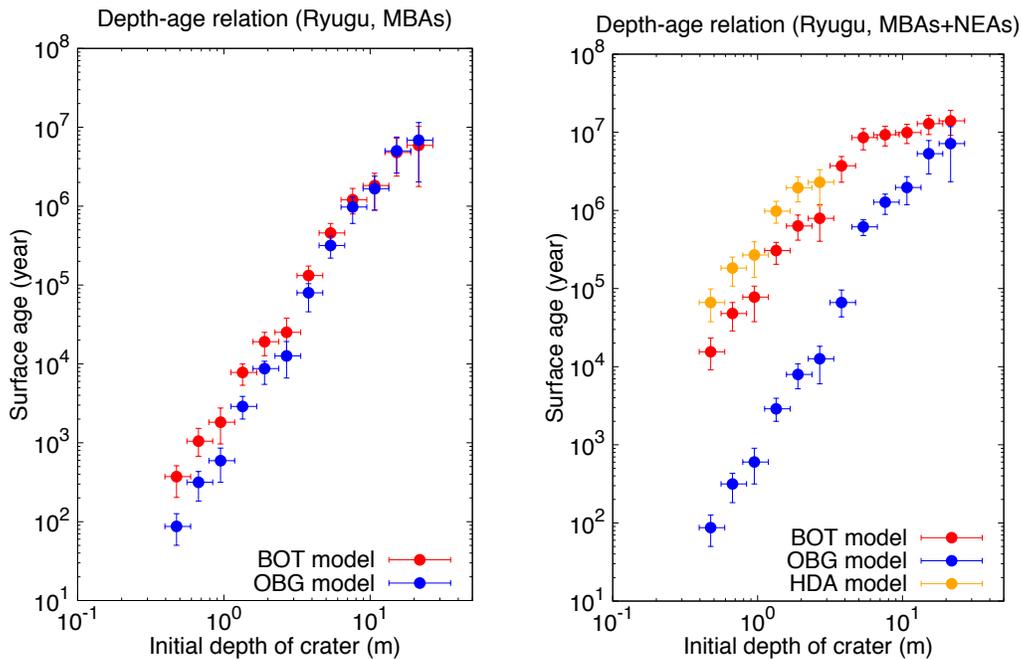

**Figure 9.** Depth-age relations of Ryugu calculated based on the population of (a) only MBAs and (b) MBAs and NEAs.

## 5 Conclusions

We estimated the retention ages (lifetime) of craters on four asteroids (Ryugu, Itokawa, Bennu, Eros) and determined that the retention age $t$ is expressed as a power-law function of the crater diameter $D$, via $t \propto D^a$. If the resurfacing is controlled by a diffusion process such as seismic shaking, the power-law index $a$ is equal to 2, or $t \propto D^2$. We first verified that dry-soil cohesion conditions (0.18 MPa) yielded a power-law index of approximately 2. However, our analyses using cohesionless conditions, which are consistent with the outcome of the Hayabusa2 impact experiment, indicated that the power-law indices are systematically greater than 2 and inconsistent with that of a diffusion process. This result suggests that the seismic shaking model does not necessarily account for the distribution of craters on small asteroids. The large power-law indices suggest that the small craters on asteroids are obliterated more rapidly than expected by the seismic shaking model.

We examined the factors that determine the power-law index. The index can vary depending on the impactor size-frequency distribution and cohesion conditions. One interpretation is that the crater obliteration process differs from the diffusion process, in which the power-law coefficient is not necessarily equal to 2. Surface flows found on Ryugu, Itokawa, and Bennu could correspond to this process. The spin rate change resulting from the YORP effect would be one of the driving forces of the flows on the asteroids. In contrast, keeping a diffusion process (seismic shaking) at the origin of crater obliteration requires that the mechanical strength or particle size of the asteroid surface differs from that at a deeper subsurface, resulting in a diffusion coefficient of the subsurface layer that is larger than that of the surface. The different intercepts of the $t - D$ relations for the different asteroids may reflect the efficiency of seismic shaking because the efficiency of seismic shaking is thought to be controlled by the surface gravity of an asteroid.

Furthermore, we estimated the relationship between the initial depth of a crater and the retention age using the crater size distribution on Ryugu. The relationship showed that the materials shallower than 1 m were mixed vertically over $10^3$–$10^5$ yr as a result of impact cratering and subsequent crater obliteration processes. This young age is comparable to the particularly young CRE age of CM and CI chondrites (0.1–1 Myr) as compared to other carbonaceous chondrites or ordinary chondrites (1–100 Myr).

The young surface age of Ryugu observed by actual crater statistics may help to resolve the inconsistency between the rapid space weathering process ($10^3$–$10^4$ yr) based on experiments and the slow space weathering based on spectroscopic observations ($10^6$–$10^7$ yr <). The timescale of resurfacing at depths shallower than 1 m on Ryugu was estimated to be $10^3$–$10^5$ yr. The resurfacing timescale is comparable to or longer than the timescale of space weathering ($10^3$–$10^4$ yr) estimated by the ion irradiation experiments. Our analysis, based on the crater data of Ryugu, suggests the slow space weathering due to the resurfacing processes.

The timescale ($10^4$–$10^6$ yr) required to resurface materials at a depth of 2–4 m can be compared with the CRE ages of returned samples to constrain the distribution of impactors colliding with Ryugu. If the CRE ages of the samples from Ryugu are measured at 1–10 Myr, the retention age using the HAD's NEA model (a few Myrs) are comparable to the CRE ages, and the retention age using the other NEA models ($10^4$–$10^6$ yr) are underestimated by an order of magnitude. This underestimation will be attributed to the uncertainties of parameters used in our model, such as the single collisional probability, impactor velocity distribution, and crater scaling laws rather than that of the NEA impactor models. In fact, the NEA impactor distribution is well-constrained by observations. A CRE age of 0.1 Myr for the Ryugu samples is consistent with the depth-age relationship for Ryugu found in this study, suggesting that the small craters present on Ryugu were formed by an impactor population similar to that of the current impactor model.

**Acknowledgments, Samples, and Data**

This paper was significantly improved by the comments from Dr. Bill Bottke and an anonymous reviewer. This study was supported by KAKENHI from the Japanese Society for Promotion of Science (JSPS) (Grant Nos. JP17H01175, JP20H00194, JP19K14778, and JP20H04607), the JSPS Core-to-Core program "International Network of Planetary Sciences" and International Graduate Program for Excellence in Earth-Space Science (IGPEES) from the University of Tokyo. P.M. acknowledges support from the French space agency, CNES and from the European Union's Horizon 2020 research and innovation program under grant agreement No 870377 (project NEO-MAPP), and from Academies of Excellence: Complex systems and Space, environment, risk, and resilience, part of the IDEX JEDI of the Université Côte d'Azur.

Tanabe, N. et al., 2021. Development of image texture analysis technique for boulder distribution measurement: Applications to asteroids Ryugu and Itokawa. *Planetary and Space Science*, Vol. 204, 105249.

Tatsumi, E., Sugita, S., 2018. Cratering efficiency on coarse-grain targets: Implications for the dynamical evolution of asteroid 25143 Itokawa. *Icarus*, Vol. 300, 227-248.

Vernazza, P. et al., 2009. Solar wind as the origin of rapid reddening of asteroid surfaces. *Nature*, Vol. 458, doi:10.1038/nature07956.

Wada, K. et al., 2018. Asteroid Ryugu before the Hayabusa2 encounter. *Progress in Earth and Planetary Science*, https://doi.org/10.1186/s40645-018-0237-y

Walsh, K.J. et al., 2019. Craters, boulders and regolith of (101955) Bennu indicative of an old and dynamic surface. *Nature geoscience*, Vol. 12, 242-246.

Watanabe, S. et al., 2019. Hayabusa2 arrives at the carbonaceous asteroid 162173 Ryugu-A spinning top-shaped rubble pile. *Science*, aav8032.

Werner, S.C. et al., 2002. The near-Earth asteroid size-frequency distribution: A snapshot of the lunar impactor size-frequency distribution. *Icarus*, Vol. 156, 287-290.

Yamada, T.M. et al., 2016. Timescale of asteroid resurfacing by regolith convection resulting from the impact-induced global seismic shaking. *Icarus*, Vol. 272, 165-177.

Yu, Y et al., 2014. Numerical predictions of surface effect during the 2029 close approach of Asteroid 99942 Apophis. *Icarus*, Vol. 242, 82-96.
30